\documentclass[aps,onecolumn,showpacs,showkeywords]{revtex4}
\usepackage{amssymb}
\usepackage{amsmath}
\usepackage{amscd}
\usepackage{graphicx}
\usepackage{subfigure}
\usepackage{epstopdf}
\usepackage{float}
\usepackage{array}
\begin{document}
\title{Comparison of the sensitivity to systematic errors between non-adiabatic
non-Abelian geometric gates and their dynamical counterparts}
\author{Shi-Biao Zheng$^{1,2}$, Chui-Ping Yang$^{1,3}$, and Franco Nori$^{1,4}$}
\address{$^1$CEMS, RIKEN, Wako-shi, Saitama 351-0198, Japan\\
$^2$Department of Physics, Fuzhou University, Fuzhou 350116, China\\
$^3$Department of Physics, Hangzhou Normal University, Hangzhou 310036, China%
\\
$^4$Department of Physics, University of Michigan, Ann Arbor, MI 48109, USA}
\date{\today }

\begin{abstract}
We investigate the effects of systematic errors of the control parameters on
single-qubit gates based on non-adiabatic non-Abelian geometric holonomies
and those relying on purely dynamical evolution. It is explicitly shown that
the systematic error in the Rabi frequency of the control fields affects
these two kinds of gates in different ways. In the presence of this
systematic error, the transformation produced by the non-adiabatic
non-Abelian geometric gate is not unitary in the computational space, and
the resulting gate infidelity is larger than that with the dynamical method.
Our results provide a theoretical basis for choosing a suitable method for
implementing elementary quantum gates in physical systems, where the
systematic noises are the dominant noise source.{\\}
\end{abstract}


\vskip 0.5cm
\maketitle
\narrowtext

\section{INTRODUCTION}

A quantum system undergoing a cyclic evolution will acquire a geometric
phase, in addition to a dynamical phase [1-6]. In contrast with the
dynamical phase, the geometric phase is only determined by some geometric
features of the closed circuit that the system traces in the projective
Hilbert space, independent of the rate at which the circuit is followed. In
the case of a cyclic adiabatic evolution of parameters in the Hamiltonian,
this additional phase is the well known Berry phase [1], which is
proportional to the area spanned in parameter space. When the subspace of
the Hilbert space traversed by the system is nondegenerate, the geometric
phase is a real number. On the other hand, the cyclic evolution of a
degenerate subspace results in a matrix-valued transformation, known as the
non-Abelian holonomy [5,6]. In addition to its fundamental interest, the
geometric phase is considered to be a candidate for the coherent control of
quantum systems. Quantum gates based on Berry phases have been investigated
both theoretically and experimentally [7-11].

The use of Berry phases for implementing fault-tolerant quantum computation
has stimulated interest in their behavior in the presence of noise [12-18].
In particular, the effect of the fluctuation noise in the classical control
parameters on the Berry adiabatic phase of a spin-1/2 system has been
analyzed [12]. It was shown [12] that the contribution of the Berry phase to
both the phase variance and dephasing vanishes for a long evolution time,
because the effect of noise on the global features of the evolution path is
averaged out in the adiabatic limit. This has been demonstrated
experimentally using ultracold neutrons [17] and circuit quantum
electrodynamics [18]. The main obstacle for the implementation of adiabatic
geometric phase gates is that the timescale associated with the adiabatic
evolution should be much longer than the dynamical one, which implies that
these geometric gates operate very slowly, compared to the dynamical
process. This makes the system vulnerable to open-system effects, which
would result in the loss of coherence. To overcome this problem, schemes
based on nonadiabatic Abelian geometric phases have been proposed [19,20],
whose sensitivity to noise has also been analyzed [21-23].

Recently, a method has been proposed for implementing holonomic quantum
computation based on non-adiabatic non-Abelian geometric phases [24]. To
realize a single-qubit gate, the two computational basis states of a qubit
are coupled to an auxiliary state by using two classical fields, which drive
the degenerate subspace spanned by the two basis states to undergo a cyclic
evolution. This kind of geometric gates have received considerable interest,
and have been experimentally demonstrated in circuit QED [25], NMR systems
[26], and solid-state spins associated with a diamond nitrogen-vacancy (NV)
center [27,28]. The effects of decoherence and noise on this gate have also
been investigated [29]. However, an open problem remains whether this gate
is more robust against errors in control parameters than the dynamical ones.
This is the aim of the present work. In the manuscript we compare the
fidelities of quantum logic gates relying on non-adiabatic non-Abelian
holonomies and the corresponding dynamical gates in the presence of
systematic errors. Here the systematic errors refer to the errors of the
corresponding control parameters that fluctuate slowly compared with the
system evolution, so that they approximately keep constant during the gate
operation. Our results show that the systematic error in the Rabi frequency
of the driving fields affects the non-adiabatic non-Abelian geometric gate
and the dynamical one in different ways: For the former, this error results
in the leakage of the qubit's population into the noncomputational space,
while for the latter it causes a deviation of the transition probability
from the expected value in the computational space. As a result, the
non-adiabatic non-Abelian geometric gate is more sensitive to the Rabi
frequency systematic error than the dynamical gate; in particular, for the
geometric implementation of a Hadamard gate, the infidelity induced by this
systematic error is increased by one order of magnitude, compared to the
dynamical method. Furthermore, the imperfect control of the amplitude ratio
of the two driving fields for realizing the non-Abelian holonomies
introduces an additional error.

This paper is organized as follows. In Sec. II, we analyze the effect of the
systematic errors of the control fields on the non-adiabatic non-Abelian
geometric gate. The result shows that the qubit has a probability of leaking
out of the computational space in the presence of the Rabi frequency
fluctuation. In Sec. III, we calculate the infidelity of the dynamical gate
caused by the systematic errors. Since the qubit remains in the
computational space, the relation between the infidelity and the parameters
of the expected gate transformation matrix is different from that for the
non-adiabatic non-Abelian geometric gate. In Sec. IV, we compare the
infidelities of the two gates, demonstrating that the geometric gate is more
sensitive to the Rabi frequency systematic error than the dynamical one.
Conclusions appear in Sec. V.

\section{FIDELITY OF NON-ADIABATIC NON-ABELIAN GEOMETRIC GATES}

We first consider the infidelity of the non-adiabatic non-Abelian geometric
operation in the presence of systematic errors in the control parameters.
Recently, it has been shown [24] that a universal holonomic single-qubit
gate based on the non-adiabatic non-Abelian geometric transformation can be
realized in a three-level system, with the two lower states, $\left|
0\right\rangle $ and $\left| 1\right\rangle $, representing the logic states
and one higher state $\left| e\right\rangle $ acting as the auxiliary state.
The transitions $\left| 0\right\rangle \rightarrow \left| e\right\rangle $
and $\left| 1\right\rangle \rightarrow \left| e\right\rangle $ are
resonantly driven by two classical fields, as shown Fig. 1(a). The
Hamiltonian in the interaction picture is given by
\begin{equation}
H_g=\hbar \Omega \left( a\left| e\right\rangle \left\langle 0\right|
+b\left| e\right\rangle \left\langle 1\right| +h.c.\right) ,
\end{equation}
where $\left| a\right| ^2+$ $\left| b\right| ^2=1$, $\Omega $ is the Rabi
frequency characterizing the transition between the bright state, $\left|
\phi _b\right\rangle =a^{*}\left| 0\right\rangle +b^{*}\left| 1\right\rangle
$, and the excited state $\left| e\right\rangle $, and $h.c.$ represents the
Hermitian conjugate. The three eigenstates of the system are
\begin{eqnarray}
\left| \phi _d\right\rangle  &=&\left( b\left| 0\right\rangle -a\left|
1\right\rangle \right) ,  \nonumber \\
\left| \phi _{b,\pm }\right\rangle  &=&\frac 1{\sqrt{2}}\left( \left| \phi
_b\right\rangle \pm \left| e\right\rangle \right) ,
\end{eqnarray}
with the corresponding eigenenergies given by
\begin{eqnarray}
E_d &=&0,  \nonumber \\
E_{b,\pm } &=&\pm \Omega .
\end{eqnarray}
The dark state $\left| \phi _d\right\rangle $ is decoupled from the
Hamiltonian and undergoes no transition during the application of the
driving fields. Without loss of generality, we here set $a=\sin (\theta
/2)e^{i\phi }$ and $b=\cos (\theta /2)$, with $0\leq \theta \leq \pi $
decided by the ratio between the amplitudes of the two driving fields and $%
0\leq \phi \leq 2\pi $ depending on the relative phase of these fields. The
evolution of the initial basis states $\left| k\right\rangle $ ($k=0,1$) are
given by $\left| \psi _k(t)\right\rangle =\exp \left( -i\int_0^tHdt/\hbar
\right) \left| k\right\rangle $. When $a/b$ remains unchanged during the
interaction, the evolution satisfies the parallel-transport condition $%
\left\langle \psi _k(t)\right| H\left| \psi _j(t\right\rangle =0$, and hence
is purely geometric. For $\int_0^T\Omega dt=\pi $, the degenerate qubit
space undergoes a cyclic evolution (the system returns to the subspace
spanned by qubit logic states $\left| 0\right\rangle $ and $\left|
1\right\rangle $), leading to a non-Abelian holonomic transformation. As a
consequence, the final evolution operator in the computational basis \{$%
\left| 0\right\rangle ,\left| 1\right\rangle $\} is

\begin{equation}
U_g=\left(
\begin{array}{cc}
\cos \theta & -\sin \theta e^{-i\phi } \\
-\sin \theta e^{i\phi } & -\cos \theta
\end{array}
\right) ,
\end{equation}
which can be used to realize any single-qubit rotation.

When the fluctuations in the amplitudes and phases of the driving fields are
considered, the Hamiltonian becomes
\begin{eqnarray}
H_g^{^{\prime }} &=&\hbar \Omega ^{^{\prime }}\left[ \sin (\theta ^{^{\prime
}}/2)e^{i\phi ^{^{\prime }}}\left| e\right\rangle \left\langle 0\right|
\right.  \nonumber \\
&&\ \ \ \left. +\cos (\theta ^{^{\prime }}/2)\left| e\right\rangle
\left\langle 1\right| \right] +h.c.{\bf ,}
\end{eqnarray}
where $\Omega ^{^{\prime }}=\Omega +\delta \Omega $, $\theta ^{^{\prime
}}=\theta +\delta \theta $, and $\phi ^{^{\prime }}=\phi +\delta \phi $,
with $\delta \Omega $, $\delta \theta $ and $\delta \phi $ being the
deviations of the corresponding parameters from the expected values due to
the presence of systematic errors of the driving fields. We note that $%
\delta \theta $ arises from the imperfect control of the ratio of the
amplitudes of the two fields. We here consider the case where the control
parameters fluctuate slowly as compared to the gate speed, so that $\delta
\Omega $, $\delta \theta $, and $\delta \phi $ can be regarded as constants
during the geometric operation. This is the case in experiments performed in
certain physical systems [17, 18, 30]. In terms of the basis states \{$%
\left| 0\right\rangle ,\left| 1\right\rangle $, $\left| e\right\rangle $\},
the evolution operator can be expressed as
\begin{equation}
U_g^{^{\prime }}=U_{g,0}^{^{\prime }}+U_{g,1}^{^{\prime }},
\end{equation}
where

\begin{equation}
U_{g,0}^{^{\prime }}=\left(
\begin{array}{ccc}
\cos \theta ^{^{\prime }} & -e^{-i\phi ^{^{\prime }}}\sin \theta ^{^{\prime
}} & 0 \\
-e^{i\phi ^{^{\prime }}}\sin \theta ^{^{\prime }} & -\cos \theta ^{^{\prime
}} & 0 \\
0 & 0 & 0
\end{array}
\right) ,
\end{equation}
\begin{equation}
U_{g,1}^{^{\prime }}=\left(
\begin{array}{ccc}
\left[ 1-\cos \left( \delta \Omega \text{\negthinspace }T\right) \right]
\sin ^2(\theta ^{^{\prime }}/2) & \frac 12\left[ 1-\cos (\delta \Omega
T)\right] \sin \theta ^{^{\prime }}e^{-i\phi ^{^{\prime }}} & i\sin \left(
\delta \Omega \text{\negthinspace }T\right) \sin \left( \theta ^{^{\prime
}}/2\right) e^{i\phi ^{^{\prime }}} \\
\frac 12\left[ 1-\cos \left( \delta \Omega \text{\negthinspace }T\right)
\right] \sin \theta ^{^{\prime }}e^{i\phi ^{^{\prime }}} & \left[ 1-\cos
\left( \delta \Omega T\right) \right] \cos ^2\left( \theta ^{^{\prime
}}/2\right) & i\sin \left( \delta \Omega \text{\negthinspace }T\right) \cos
\left( \theta ^{^{\prime }}/2\right) \\
i\sin (\delta \Omega \text{\negthinspace }T)\sin \left( \theta ^{^{\prime
}}/2\right) e^{-i\phi ^{^{\prime }}} & i\sin (\delta \Omega \text{%
\negthinspace }T)\cos \left( \theta ^{^{\prime }}/2\right) & -\cos (\delta
\Omega \text{\negthinspace }T)
\end{array}
\right) .
\end{equation}
This result shows that the Rabi frequency fluctuation causes the leakage of
the qubit's population out of the computational space.

For any input state $\left| \psi _i\right\rangle $, the fidelity of the
output state is defined as

\begin{equation}
{\cal F}_{g,\psi ,1}=\left| \left\langle \psi _d\right| \left. \psi
_r\right\rangle \right| ^2,
\end{equation}
where $\left| \psi _d\right\rangle =U_g\left| \psi _i\right\rangle $ is the
desired output state, and $\left| \psi _r\right\rangle =$ $U_g^{^{\prime
}}\left| \psi _i\right\rangle $ is the real output state with the amplitude
and phase errors of the driving fields being considered. Setting $\left|
\psi _i\right\rangle =\cos \frac \alpha 2\left| 0\right\rangle +\sin \frac
\alpha 2e^{i\beta }\left| 1\right\rangle $ [23, 31], to second order in $%
\delta $\negthinspace $\theta $, $\delta \phi $, and $\delta \Omega T$, we
obtain
\begin{eqnarray}
{\cal F}_{g,\psi } &\simeq &1-(\delta \theta )^2\left[ 1-\sin ^2\alpha \sin
^2(\beta -\phi )\right]  \nonumber \\
&&\ \ -(\delta \phi )^2\left\{ \sin ^2\theta -\left[ \cos \alpha \sin
^2\theta +\frac 12\sin \alpha \sin (2\theta )\cos (\beta -\phi )\right]
^2\right\}  \nonumber \\
&&\ \ -2\delta \theta \delta \phi \sin \alpha \sin (\beta -\phi )\left[ \cos
\alpha \sin ^2\theta +\frac 12\sin \alpha \sin (2\theta )\cos (\beta -\phi
)\right]  \nonumber \\
&&\ \ -(\delta \Omega T)^2\left| \cos \alpha \sin (\theta /2)e^{i\phi }+\sin
\alpha \cos (\theta /2)e^{i\beta }\right| ^2.
\end{eqnarray}
For fixed parameter errors, the fidelity depends on the initial qubit state.
Averaging over all the input states, the average fidelity is calculated to be

\begin{eqnarray}
{\cal F}_g &=&\text{$\frac 1{4\pi }\int_0^\pi $ \negthinspace }\sin \alpha
d\alpha \text{$\int_0^{2\pi }$ \negthinspace }d\beta \text{ \negthinspace $%
{\cal F}_{g,\psi }$}  \nonumber \\
\ &=&1-\frac 13\left( \pi \frac{\delta \Omega }\Omega \right) ^2\left[
1+\cos ^2(\theta /2)\right] -\frac 23(\delta \text{\negthinspace }\theta )^2
\nonumber \\
&&\ \ \ \ -(\delta \phi )^2\left[ \frac 23\sin ^2\theta -\frac 1{12}\sin
^2(2\theta )\right] .
\end{eqnarray}
We here have used the relation $\Omega $\negthinspace $T=\pi $. The result
shows that the gate infidelity arising from the Rabi frequency systematic
error (the second term of ${\cal F}_g$) decreases as the parameter $\theta $
increases.

\section{FIDELITY OF THE DYNAMICAL GATE}

Let us now consider the effects of systematic errors of the control
parameters on gates based on purely dynamical evolution, which can be
achieved by resonantly driving the transition between the two logic states $%
\left| 0\right\rangle $ and $\left| 1\right\rangle $ with a classical field,
as shown in Fig. 1(b). The corresponding Hamiltonian is given by
\begin{equation}
H_d=\hbar \left( \Omega e^{-i\phi }\left| 0\right\rangle \left\langle
1\right| +\Omega e^{i\phi }\left| 1\right\rangle \left\langle 0\right|
\right) .
\end{equation}
With this Hamiltonian one can obtain an evolution operator similar to the
geometric gate
\begin{equation}
U_d=\left(
\begin{array}{cc}
\cos \theta & -ie^{-i\phi }\sin \theta \\
-ie^{i\phi }\sin \theta & \cos \theta
\end{array}
\right) ,
\end{equation}
where $\theta =\Omega \tau $, with $\tau $ being the gate duration. Any
single-qubit transformation can be obtained through this kind of evolutions
with $0<\theta \leq \pi /2$. For example, with the choice $\phi =3\pi /2$
this evolution yields the gate $U_y(\phi _y)=\exp \left( i\phi _y\sigma
_y\right) $ with $\phi _y=\theta $. Two sequential evolutions $U_d(\theta
=\pi /2,\phi =\phi _z/2)$ and $U_d(\theta =\pi /2,\phi =-\phi _z/2)$ produce
the gate $U_z=\exp \left( i\phi _z\sigma _z\right) $ plus a trivial global $%
\pi -$phase shift. The rotations $U_y(\phi _y)$ and $U_z(\phi _z)$ with $%
0<\phi _y<\pi $ and $0<\phi _z<\pi $ form a universal set of single-qubit
operations. The same single-qubit operation can be achieved using a similar
combination of non-adiabatic non-Abelian geometric gates [27]. In this
sense, the effect of the dynamical gate $U_d$ is equivalent to that of the
geometric gate $U_g$.

With the field amplitude and phase deviations $\delta \Omega $ and $\delta
\phi $\ being included, the qubit evolution operator can now be written as
\begin{equation}
U_d^{^{\prime }}=\left(
\begin{array}{cc}
\cos \theta ^{^{\prime }} & -ie^{-i\phi ^{^{\prime }}}\sin \theta ^{^{\prime
}} \\
-ie^{i\phi ^{^{\prime }}}\sin \theta ^{^{\prime }} & \cos \theta ^{^{\prime
}}
\end{array}
\right) ,
\end{equation}
where $\theta ^{^{\prime }}=\theta +\delta \Omega \!\tau $ and $\phi
^{^{\prime }}=\phi +\delta \phi $. This result shows that the Rabi frequency
fluctuation leads to the imperfect control of the probability for the
qubit's transition between the computational states $\left| 0\right\rangle $
and $\left| 1\right\rangle $. For the input state $\left| \psi
_i\right\rangle =\cos \frac \alpha 2\left| 0\right\rangle +\sin \frac \alpha
2e^{i\beta }\left| 1\right\rangle $, to second order in $\delta \Omega
\!\tau $ and $\delta \phi $, the fidelity of the output state becomes
\begin{eqnarray}
{\cal F}_{d,\psi } &\simeq &1-(\delta \Omega \text{\negthinspace }\tau
)^2\left[ 1-\sin ^2\alpha \cos ^2(\beta -\phi )\right]  \nonumber \\
&&\ \ \ \ \ \ \ \ -(\delta \phi )^2\left\{ \sin ^2\theta -\left[ \cos \alpha
\sin ^2\theta +\frac 12\sin \alpha \sin (2\theta )\sin (\beta -\phi )\right]
^2\right\}  \nonumber \\
&&\ \ \ \ \ \ \ \ -2\delta \phi \delta \Omega \text{\negthinspace }\tau \sin
\alpha \cos (\beta -\phi )\left[ \cos \alpha \sin ^2\theta -\frac 12\sin
\alpha \sin (2\theta )\sin (\beta -\phi )\right] .
\end{eqnarray}
Averaging over all the input states and using the relation $\theta =\Omega
\!\tau $, we obtain the average fidelity for the dynamical gate

\begin{equation}
{\cal F}_d\simeq 1-\frac 23\left( \theta \frac{\delta \Omega }\Omega \right)
^2-(\delta \text{ }\!\phi )^2\left[ \frac 23\sin ^2\theta -\frac 1{12}\sin
^2(2\theta )\right] .
\end{equation}
The result shows that the gate infidelity induced by the Rabi frequency
systematic error (the second term of ${\cal F}_d$) increases quadratically
with $\theta $, which is in distinct contrast with the case for the
non-adiabatic non-Abelian geometric gate.

\section{COMPARISON}

The error of the dynamical gate due to the Rabi frequency fluctuation $%
\delta \Omega $ is proportional to the square of the parameter $\theta $ of
the expected gate transformation, which is in stark contrast with the
non-adiabatic non-Abelian geometric gate, whose infidelity associated with $%
\delta \Omega $ is proportional to $1+\cos ^2(\theta /2)$. This difference
is due to the fact that the required operation time $\tau =\theta /\Omega $
is proportional to $\theta $ for the dynamical evolution, while the
geometric operation time does not depend upon $\theta $ for a given Rabi
frequency $\Omega $. In other words, the condition $\Omega \!T=\pi $ is
always required to implement any non-adiabatic non-Abelian geometric gate.
As a result, the error of the dynamical gate shows a much stronger
dependence on $\theta $ than that of the geometric gate. We note that the
fluctuation $\delta \Omega $ affects the non-adiabatic non-Abelian geometric
gate and the dynamical one in different ways. For the non-adiabatic
non-Abelian geometric gate, this fluctuation leads to the leakage of the
population from the bright state $\left| \phi _b\right\rangle $ to the
auxiliary state $\left| e\right\rangle $ after the operation, so that the
resulting transformation in the computational space $\left\{ \left|
0\right\rangle ,\left| 1\right\rangle \right\} $ is not unitary. In
contrast, for the dynamical method this fluctuation results in inaccurate
control of the transition probability between the two logic states $\left|
0\right\rangle $ and $\left| 1\right\rangle $; these two states span the
whole state space and the resulting transformation in the computational
space remains unitary. It should be noted that the second term of Eq. (16)
can be expressed as $-2(\delta \theta )^2/3$, which has the same form as the
third term of Eq. (11). However, the reasons for the occurence of the error $%
\delta \theta $ are completely different for the non-adiabatic non-Abelian
geometric gate and the dynamical one. For the former, this error is caused
by imperfect control of the ratio between the amplitudes of the two fields
driving the transitions $\left| 0\right\rangle \rightarrow \left|
e\right\rangle $ and $\left| 1\right\rangle \rightarrow \left|
e\right\rangle $, and is equal to $\delta \theta =2\frac{\delta r}{1+r^2}$,
where $r=\left| a/b\right| $, with $\left| a\right| \Omega $ and $\left|
b\right| \Omega $ characterizing the amplitudes of the two driving fields,
as shown in Eq. (1). For the latter, $\delta \theta $ is due to the Rabi
frequency fluctuation, and is given by $\delta \theta =\delta \Omega \tau $.

Since $\theta \leq \pi /2$, the maximum infidelity induced by the error $%
\delta \Omega $ for the dynamical operation $U_d$ is $\frac 16\left( \pi
\frac{\delta \Omega }\Omega \right) ^2$, which is only one half of the
minimum of the corresponding infidelity for the non-adiabatic non-Abelian
geometric operation $U_g$, when the values of the relative error $\delta
\Omega /\Omega $ are the same for both gates. More importantly, for the
dynamical method this infidelity decreases quadratically as the parameter $%
\theta $ decreases. For example, for the dynamical implementation of the
Hadamard transformation this infidelity is reduced by one order of
magnitude, compared to the geometric method. Since only one field is needed
for realizing $U_d$, the error associated with the fluctuation in the ratio
between the amplitudes of the two fields used for implementing $U_g$ does
not appear in the dynamical method. For the same $\theta $ and $\delta
\!\phi $, the errors due to the field phase fluctuation are the same for
both cases. Therefore, when the systematic errors are the main error source,
the fidelity of the non-Abelian non-adiabatic geometric operation is lower
than the corresponding dynamical one.

We further note that the required dynamical gate time $\theta /\Omega $ is
shorter than $\pi /(2\Omega )$ as $\theta \leq \pi /2$, while the geometric
gate duration is always $\pi /\Omega $. Due to the longer operation time,
the non-adiabatic non-Abelian geometric gate is more sensitive to
decoherence effects, arising from the system-reservoir coupling, than the
dynamical gate.

\section{CONCLUSION}

In conclusion, we have analyzed the effects of systematic errors in the
control parameters on the non-adiabatic non-Abelian geometric gate and on
its dynamical equivalent. We show that the systematic error of the Rabi
frequency of the driving fields affects these two gates in different ways.
This systematic error causes the qubit's population to leak to the
noncomputational space after the non-adiabatic non-Abelian geometric gate is
performed, while the qubit always remains in the computational space when
the gate is implemented by the dynamical method. As a consequence, the
non-adiabatic non-Abelian geometric gates are more sensitive to the Rabi
frequency systematic error than the purely dynamical ones. For
implementation of the Hadmarda gate, the infidelity induced by this error in
the non-adiabatic non-Abelian geometric method is one order of magnitude
larger than that in the dynamical one. Our results are usful for choice of
optimal methods for implementing elementary quantum gates under the
condition that the fluctuations of the control parameters are slow compared
with the gate speed.

{\bf ACKNOWLEDGEMENTS}

SBZ acknowledges support from the Major State Basic Research Development
Program of China under Grant No. 2012CB921601 and the National Natural
Science Foundation of China under Grant No. 11374054. CPY acknowledges
support from the National Natural Science Foundation of China under Grant
No. 11374083. FN is partially supported by the RIKEN iTHES Project, the MURI
Center for Dynamic Magneto-Optics via the AFOSR award number
FA9550-14-1-0040, the IMPACT program of JST, and a Grant-in-Aid for
Scientific Research (A)

\begin{figure}
\center
  \includegraphics[width=0.5\columnwidth]{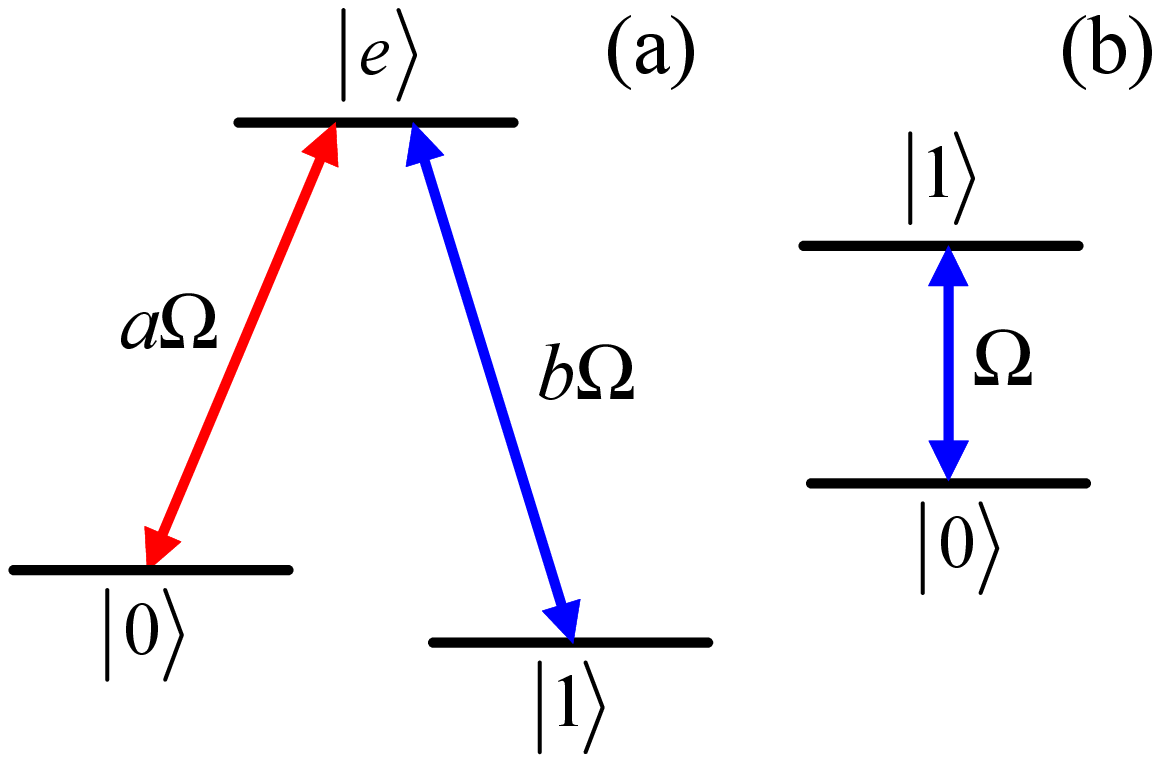}\caption{(color online). 
The system level configuration and excitation scheme.
(a). Non-Abelian non-adiabatic gate: The qubit is represented by two ground
states $\left| 0\right\rangle $ and $\left| 1\right\rangle $ of a $\Lambda -$%
type three-level system, with the transitions $\left| 0\right\rangle
\rightarrow \left| e\right\rangle $ and $\left| 1\right\rangle \rightarrow
\left| e\right\rangle $ being resonantly driven by two classical fields with
the couplings $a\Omega $ and $b\Omega $, respectively, with $\left| a\right|
^2+$ $\left| b\right| ^2=1$. $\Omega $ is the Rabi frequency characterizing
the transition between the bright state $\left| \phi _{b,0}\right\rangle
=a^{*}\left| 0\right\rangle +b^{*}\left| 1\right\rangle $ and the auxiliary
excited state $\left| e\right\rangle $. (b) Dynamical gate: The transition $%
\left| 0\right\rangle \rightarrow \left| 1\right\rangle $ is directly
coupled by a classical field with Rabi frequency $\Omega $.}\label{fig1}
\end{figure}


\begin{references}
\bibitem{}  M. V. Berry, Quantal phase-factors accompanying adia- batic
changes, {\it Proc. R. Soc. Lond. }A {\bf 392}, 45 (1984).

\bibitem{}  J. Anandan, The geometric phase, {\it Nature} {\bf 360}, 307
(1992).

\bibitem{}  Y. Aharonov and J. Anandan, Phase-change during a cyclic
quantumevolution, {\it Phys. Rev. Lett.} {\bf 58}, 1593 (1987).

\bibitem{}  S. Oh, X. Hu, F. Nori, and S. Kais, Singularity of the
time-energy uncertainty in adiabatic perturbation and cycloids on a Bloch
sphere, Sci. Rep. 6, 20824 (2016).

\bibitem{}  F. Wilczek and A. Zee, Appearance of gauge structure in simple
dynamical systems, {\it Phys. Rev. Lett. }{\bf 52}, 2111 (1984).

\bibitem{}  J. Anandan, Non-adiabatic non-Abelian geometric phase, {\it %
Phys. Lett.} A {\bf 133}, 171 (1988).

\bibitem{}  P. Zanardi and M. Rasetti, Holonomic quantum computation, {\it %
Phys. Lett.} A {\bf 264}, 94 (1999).

\bibitem{}  G. Falci, R. Fazio, G. M. Palma, J. Siewert, and V. Vedral,
Detection of geometric phases in superconducting nanocircuits, {\it Nature}
{\bf 407}, 355 (2000).

\bibitem{}  L. M. Duan, J. I. Cirac, and P. Zoller, Geometric manipulation
of trapped ions for quantum computation, {\it Science }{\bf 292}, 1695
(2001).

\bibitem{}  S. B.\ Zheng, Unconventional geometric quantum phase gates with
a cavity QED system, {\it Phys. Rev.} A {\bf 70}, 052320 (2004).

\bibitem{}  J. A. Jones, V. Vedral, A. Ekert, and G. Castagnoli, Geometric
quantum computation with NMR, {\it Nature} {\bf 403}, 869 (2000).

\bibitem{}  G. De Chiara and G. M. Palma, Berry phase for a spin-1/2
particle in a classical fluctuating field, {\it Phys. Rev. Lett.} {\bf 91},
090404 (2003).

\bibitem{}  A. Carollo, I. Fuentes-Guridi, M. F. Santos, and V. Vedral,
Geometric phase in open systems, {\it Phys. Rev. Lett.} {\bf 90}, 160402
(2003).

\bibitem{}  A. Carollo, I. Fuentes-Guridi, M. F. Santos, and V. Vedral,
Spin-1/2 geometric phase driven by decohering quantum fields, {\it Phys.
Rev. Lett.} {\bf 92}, 020402 (2004).

\bibitem{}  S. B. Zheng, Manifestation of a nonclassical Berry phase of an
electromagnetic field in atomic Ramsey interference, {\it Phys. Rev.} A {\bf %
85}, 022128 (2012).

\bibitem{}  P. J. Leek, J. M. Fink, A. Blais, R. Bianchetti, M. G\"oppl, J.
M. Gambetta, D. I. Schuster, L. Frunzio, R. J. Schoelkopf, A. Wallraff,
Observation of Berry's Phase in a Solid State Qubit, {\it Science} {\bf 318}%
, 1889 (2007).

\bibitem{}  S. Filipp, J. Klepp, Y. Hasegawa, C. Plonka-Spehr, U. Schmidt,
P. Geltenbort, and H. Rauch, Experimental demonstration of the stability of
Berry's phase for a spin-1/2 particle, {\it Phys. Rev. Lett.} {\bf 102},
030404 (2009).

\bibitem{}  S. Berger, M. Pechal, A. A. Abdumalikov, C. Eichler, L. Steffen,
A. Fedorov, A. Wallraff, and S. Filipp, Exploring the Effect of Noise on the
Berry Phase, {\it Phys. Rev.} A {\bf 87}, 060303(R) (2013).

\bibitem{}  X. B. Wang and M. Keiji, Nonadiabatic Conditional Geometric
Phase Shift with NMR, {\it Phys. Rev. Lett.} {\bf 87}, 097901 (2001).

\bibitem{}  S. L. Zhu and Z. D. Wang, Implementation of universal quantum
gates based on nonadiabatic geometric phases, {\it Phys. Rev. Lett.} {\bf 89}%
, 097902 (2002).

\bibitem{}  A. Nazir, T. Spiller, and W. J. Munro, Decoherence of geometric
phase gates, {\it Phys. Rev.} A {\bf 65}, 042303 (2002).

\bibitem{}  A. Blais and A.-M. S. Tremblay, Effect of noise on geometric
logic gates for quantum computation, {\it Phys. Rev.} A {\bf 67}, 012308
(2003).

\bibitem{}  J. T. Thomas, M. Lababidi, and M. Tian, Robustness of
single-qubit geometric gate against systematic error, {\it Phys. Rev.} A
{\bf 84}, 042335 (2011)

\bibitem{}  E. Sj\"oqvist, D. M. Tong, L. M. Andersson, B. Hessmo, M.
Johansson, and K. Singh, Non-adiabatic holonomic quantum computation, {\it %
New J. Phys. }{\bf 14}, 103035 (2012).

\bibitem{}  A. A. Abdumalikov Jr, J. M. Fink, K. Juliusson, M. Pechal, S.
Berger, A. Wallraff, and S. Filipp, Experimental realization of non-Abelian
non-adiabatic geometric gates, {\it Nature} {\bf 496}, 482 (2013).

\bibitem{}  G. Feng, G. Xu, and G. Long, Experimental realization of
nonadiabatic holonomic quantum computation, {\it Phys. Rev. Lett.} {\bf 110}%
, 190501 (2013).

\bibitem{}  C. Zu, W.-B. Wang, L. He, W.-G. Zhang, C.-Y. Dai, F. Wang, and
L.-M. Duan, Experimental realization of universal geometric quantum gates
with solid-state spins, {\it Nature} {\bf 514}, 72 (2014).

\bibitem{}  S. Arroyo-Camejo, A. Lazariev, S. W. Hell, and G.
Balasubramanian, Room temperature high-fidelity holonomic single-qubit gate
on a solid-state spin, {\it Nature Comm.} 5, 4870 (2014).

\bibitem{}  M. Johansson, E. Sj\"oqvist, L. M. Andersson, M. Ericsson, B.
Hessmo, K. Singh, and D. M. Tong, Robustness of non-adiabatic holonomic
gates, {\it Phys. Rev. }A {\bf 86}, 062322 (2012).

\bibitem{}  J. Benhelm, G. Kirchmair, C. F. Roos, and R. Blatt, Towards
fault-tolerant quantum computing with trapped ions, {\it Nature Phys}. {\bf 4%
}, 463 (2008).

\bibitem{}  M. A. Nielsen and I. L. Chuang, Quantum Computation and Quantum
Information (Cambridge Univ. Press, 2000).
\end{references}
\end{document}